\documentclass{article}

\usepackage[a4paper,
    left=25mm,
    right=25mm,
    top=25mm,
    bottom=25mm]{geometry}

\usepackage[T1]{fontenc}
\usepackage{cite}
\usepackage{graphicx}

\usepackage{amsthm}
\usepackage{amssymb}
\usepackage{amsmath}
\usepackage{mathtools}
\usepackage[hidelinks,colorlinks=true,linkcolor=blue,citecolor=blue]{hyperref}
\usepackage[capitalise,nameinlink]{cleveref}

\usepackage{paralist}
\usepackage{amsfonts}
\usepackage{subcaption}
\usepackage{makecell}

\usepackage{tikz}
\usetikzlibrary{positioning, calc, fit, math, trees}
\usetikzlibrary{decorations.pathreplacing,calligraphy}

\usepackage{todonotes}
\usepackage{wrapfig}
\usepackage{orcidlink}
\usepackage[norefs,nocites,nomsgs]{refcheck}

\usepackage{graphicx}

\newcommand{\mathttt}[1]{\textup{\texttt{#1}}}

\newcommand{\bigO}{\mathcal{O}}
\newcommand{\Oh}[1]{\bigO\left( #1 \right)}
\newcommand{\Ohs}[1]{\bigO( #1 )}

\newcommand{\dd}{\mathinner{.\,.}}
\newcommand{\emptystring}{\varepsilon}

\newcommand{\Arr}{{\mathsf A}}
\newcommand{\PPD}{\mathrm{PRD}}
\newcommand{\sep}{,\,}
\newcommand{\PPDc}{(i\sep|q_0|\sep|q_1|\sep k\sep e)}

\newcommand{\Pal}{\mathcal{P}}
\newcommand{\Run}{\mathcal{R}}
\renewcommand{\root}{r}

\newcommand{\Runi}[1]{\Run^{(#1)}}
\newcommand{\rooti}[1]{\root^{(#1)}}

\newcommand{\Runc}[1]{\Run_{c_{#1}}}
\newcommand{\rootc}[1]{\root_{c_{#1}}}
\newcommand{\Palc}[1]{\Pal_{c_{#1}}}

\newcommand{\abs}[1]{\vert #1 \vert}
\newcommand{\len}[1]{\vert #1 \vert}

\newcommand{\I}{\mathbf I}

\newcommand{\apc}{\hat{c}_1}
\newcommand{\shr}[2]{\operatorname{shr}_{#2}(#1)}
\newcommand{\shl}[2]{\operatorname{shl}_{#2}(#1)}
\newcommand{\scl}[2]{\operatorname{scl}_{#2}(#1)}

\newtheorem{theorem}{Theorem}[section]
\newtheorem{corollary}[theorem]{Corollary}
\newtheorem{lemma}[theorem]{Lemma}
\newtheorem{example}[theorem]{Example}

\newtheorem{observation}[theorem]{Observation}

\newtheorem{definition}[theorem]{Definition}
\newtheorem{remark}[theorem]{Remark}

\begin{document}
\bibliographystyle{alpha}

\title{Asymptotically Optimal Representation of Palindromic Structure}
\date{}
\author{Michael Itzhaki\orcidlink{0009-0009-4783-2537}}

\maketitle

\begin{abstract}
We introduce an asymptotically optimal representation of the Manacher array of a string that supports constant-time access. The approach relies on the combinatorial properties of palindromes, yielding a compact yet efficient structure.
This work fits within the broader study of compressed text indexing and highlights structural aspects of palindromic substrings that may inspire further algorithmic applications.
\end{abstract}

\section{Introduction}\label{s:intro}

Palindromes are sequences of characters that read the same forward and backward and have fascinated mathematicians and linguists for centuries due to their unique properties. Since the inception of modern stringology, identifying palindromes in a string has been an active field of research~\cite{man:75, Galil1976, KMP-77}. Palindromes represent a simple yet profound form of symmetry found not only in language but also in number theory~\cite{LSD2021:antipalindrome} and DNA sequences~\cite{sr:12}, and have applications ranging from error detection in coding theory~\cite{Lenz2019} to the study of self-replicating structures in bioinformatics~\cite{lssnz:05,gk:97,f:03}.

The \emph{Manacher array} is a fundamental structure for palindrome processing.
For each center $c$ (integer or half-integer), the Manacher array stores the radius of the longest palindrome mirrored around $c$. The array can be computed in linear time~\cite{man:75, Galil1976} and has become pervasive in string algorithms~\cite{Jeuring1994, Rubinchik2018, Charalampopoulos2022, Kosolobov2015}.

Storing the Manacher array naively requires $\Ohs{n\log n}$ bits, even though $\Theta(n)$ bits are necessary (folklore; for example, see~\cite{Itzhaki2025}). 
This raises a natural and previously unresolved question:  
\emph{Can the Manacher array be encoded in $\Ohs{n}$ bits while still supporting efficient access?}

Classical compression algorithms, such as Huffman coding~\cite{H-52}, the Lempel-Ziv family~\cite{LZ-76,LZ-85,MW-85,ZL-77}, and succinct data structures like bitvectors~\cite{Jacobson1989,Clark1997}, wavelet trees~\cite{GGV03,Navarro2012}, and FIDs~\cite{Raman2007}, provide a general framework for compact sequence representation. However, these do not exploit the specific combinatorial structure of palindromes.

In this work, we study the compression of the Manacher array in the setting of compressed pattern matching, a line of research that has received significant recent attention~\cite{Kempa2023FOCS, Ganardi2022, Gawrychowski2024, Kempa2023SODA}. 
Whereas the cited prior work has focused on compressing the strings or patterns themselves, we instead target a derived combinatorial structure, aiming to compress it without sacrificing access efficiency. Our specific line of research also received attention; a preliminary version of this work appeared as a poster in DCC 2025~\cite{Itzhaki2025Poster} and a recent preprint~\cite{mieno2025palindromes} studies a related problem and obtains results of a similar flavor.

\paragraph{Our contributions.} This paper studies the compression of the Manacher array of a string; we show a compact representation of the Manacher array that supports constant-time access. Our contribution is summarized in the following theorem:

\begin{theorem}\label{th:main}
    Let $S$ be a string of length $n$, and let $\mathsf{A}$ denote its Manacher array. The array $\mathsf{A}$ admits a lossless encoding of size $\Ohs{n}$ bits that supports element access in constant time.
\end{theorem}

This paper advances the well-established line of research on the combinatorial properties of palindromes, providing refined structural insights and extending the current understanding of their behavior~\cite{Itzhaki2025, tomohiro:10, Kosolobov2015, Nagashita2023, RS20:pal, AI24:pal}.

\section{Preliminaries}\label{s:pre}

\paragraph{Computation model.} We assume the Word RAM model with a word length of size $\Omega(\log n)$. 
All indices in this paper are 1-based, and logarithms are taken to base two. In this model, logarithms can be computed in constant time.

\paragraph{Basic definitions.} A {\em string} is an ordered sequence of characters. We denote the length of $S$ as $|S|$, its $i$-th character by $S[i]$ and its factor $S[i]S[i+1]\dots S[j]$ by $S[i\dd j]$. The empty string of length $0$ is denoted by $\emptystring$. When $j<i$, we denote $S[i\dd j]$ as the empty string $\emptystring$. A factor $S[1\dd i]$ is called a {\em prefix}, and a factor $S[i\dd |S|]$ is called a {\em suffix}. If $1\le i<|S|$ (resp. $|S|\ge j>1$) the prefix (resp. suffix) is said to be {\em proper}. The {\em concatenation} of strings $S$ and $T$ is $S\cdot T\coloneq S[1]S[2]\dots S[|S|]T[1]T[2]\dots T[|T|]$. When clear from context, we simply write $ST$. For a positive integer $i$, we denote $S^i$ as the concatenation of $S$ to itself $i$ times, i.e., $S^1 = S$, and $S^i = S^{i-1} \cdot S$.
To refer to a specific occurrence of a factor in a string, we use the interval notation $[i\dd j]=\{t\mid t\in \mathbb N,\; i \le t \le j\}$. An interval $[i\dd t]$ is called a {\em prefix} of $[i\dd j]$ if $t\le j$, and a suffix is defined symmetrically.
The center of an interval $\mathcal P=[i\dd j]$ is denoted as $c_\Pal \coloneq \frac{i+j}{2}$.

\paragraph{Periods and repetitions.} A string $S$ has {\em period $p$} if $p$ is a positive integer and $S[i]=S[i+p]$ for every $1\le i\le |S|-p$. The {\em period} of $S$ is the smallest such value $p$, and is denoted by $\operatorname{period}(S)$. A string $S$ is said to be {\em periodic} if $\operatorname{period}(S)\le |S|/2$. A {\em repetition} in $S$ is an interval $[i\dd j]\subseteq [1\dd |S|]$ for which the corresponding factor $S[i\dd j]$ is periodic. If $p=\operatorname{period}(S)$, we refer to the factor $S[1\dd p]$ as the {\em root} of $S$. The root of a repetition is defined as the root of its corresponding factor.

A {\em run} (or {\em maximal repetition}) in a string is a repetition that cannot be extended to the left or to the right with the same period. For instance, consider $S=\mathttt{cababab}$. The repetition $[3\dd 7]$ corresponds to the factor $\mathttt{babab}$ with period 2. However, it is not a run, as it can be extended by one character to the left, resulting in the repetition $[2\dd 7]$ which has the same period. In contrast, the repetition $[2\dd 7]$ is a run. Note that any repetition in a string can be extended (possibly by zero characters) to a unique run of the same period. We refer to the result of the extension as the {\em periodic extension} of the repetition.

\paragraph{Palindromes.} A string $S$ is called a \textit{palindrome} or \textit{palindromic} if $S$ equals its reversal, i.e., $S[i]=S[|S|-i+1]$ for every $1 \le i \le |S|$. For example, the strings $\mathttt{abcba}$ and $\mathttt{abccba}$ are palindromes, while $\mathttt{abcbaa}$ is not. An interval $\Pal=[i\dd j]$ is called a {\em subpalindrome} of $S$ if its corresponding factor $S[i\dd j]$ is palindromic. We say that $\Pal$ is {\em centered} at $c_{\Pal}$, and that its {\em radius} is $\lceil \frac{j-i}{2}\rceil$. A palindromic prefix (resp. suffix) is the subpalindrome $[1\dd i]$ (resp. $[i\dd |S|]$). The interval $\Pal$ is called a {\em maximal subpalindrome} if $\Pal$ is a subpalindrome and $S[i-1\dd j+1]$ is either not defined or not a palindrome. We regard the empty prefix $S[1 \dd 0]$ and empty suffix $S[|S|+1 \dd |S|]$ as subpalindromes of radius $0$ centered at $0.5$ and $|S|+0.5$, respectively.
For every center $c \in \{1, 1.5, \dots, |S|\}$ there is a corresponding unique maximal subpalindrome $\Pal$ with $c_\Pal = c$. The Manacher array records for every center the radius of its corresponding maximal subpalindrome.
\begin{definition}[Manacher array]
Let $S$ be a string of length $n$. The \emph{Manacher array} of $S$ is the array $\mathsf{A}[1 \dd  2n+1]$ where each position $i$ corresponds to a center of $S$, and is defined as follows:
\begin{itemize}
    \item If $i$ is even, $i = 2k$, then $\mathsf{A}[i]$ is the maximum integer $r \geq 0$ such that $S[k - r \dd k + r]$ is a palindrome, and $1 \le k - r \le  k + r \le n$.
    \item If $i$ is odd, $i = 2k+1$, then $\mathsf{A}[i]$ is the maximum integer $r \geq 0$ such that $S[k-r+1\dd k + r]$ is a palindrome, and $1 \le k - r + 1,\; k + r \le n$.
\end{itemize}
\end{definition}

The following lemma correlates periodicity with palindromes:
\begin{lemma}[Folklore, \cite{Inenaga2014}]\label{p:pal-per}
Let $P$ be a palindrome. If the palindrome $P$ has a proper palindromic suffix of length $p$ then $|P|-p$ is a period of $P$.
\end{lemma}

\paragraph{Additional notation.} When $S$ is clear from context, we denote the maximal subpalindrome of $S$ at center $c$ as $\Pal_c$. When $\Pal_c$ is periodic, we denote by $\Run_c$ its periodic extension and by $\root_c$ the root of $\Run_c$. Notice that $\Pal_c$ and $\Run_c$ are intervals, while $\root_c$ is a string. Additionally, note that the root of the factor corresponding to $\Pal_c$ is not necessarily $\root_c$.

\paragraph{Bitvector operations.} We use the standard \emph{rank} and \emph{select} operations on bitvectors. Given $v \in \{0,1\}^n$, $\operatorname{rank}_1(v,i)$ returns the number of 1-bits in $v[1\dd i]$, and $\operatorname{select}_1(v,i)$ returns the position of the $i$-th 1 in $v$. 
The bitvector can be preprocessed into a structure of size $n + o(n)$ bits that supports both operations in constant time~\cite{Jacobson1989,Clark1997}.

Standard bitwise shift operations scale indices by powers of 2. To adapt these to 1-based indexing, for an index $i$ and an integer $m \ge 0$, we define the 1-based right shift as $\shr{i}{m} \coloneq ((i-1) \gg m) + 1$, and the 1-based left shift as $\shl{i}{m} \coloneq ((i-1) \ll m) + 1$. Recall that $i\ll m=i\cdot 2^m$ and $i\gg m=\lfloor i\cdot 2^{-m}\rfloor$.

\paragraph{SDC and Smooth arrays.}

The \emph{Simple Dense Coding} (SDC) representation~\cite{fn1} binary-encodes all elements in an array $\Arr$ of non-negative integers, concatenates the codes, and marks the beginning of each codeword in an auxiliary bitvector. Supporting {\em select} on the auxiliary bitvector yields constant-time access to any element, while requiring $\Oh{\sum_{i=1}^{n} \log (\Arr[i]+1)}$ bits of space. We call the latter expression the {\em log-sum} of $\Arr$. In particular, an array of non-negative integers that admits a linear log-sum can be SDC-encoded using $\Oh{n}$.
\begin{remark}
    In general, Manacher arrays do not admit a linear log-sum.
\end{remark}
For instance, consider the string $S=a^n$. The log-sum of its Manacher array is $\Oh{n\log n}$, hence the SDC-encoding of a Manacher array requires $\omega(n)$ bits of space in the worst case.

Our ultimate goal is to encode a Manacher array with $2n+1$ elements using $\Oh{n}$ bits while still supporting constant-time access. To do so, we construct arrays with linear log-sum and encode them using SDC. To show that an array has linear log-sum we impose a stronger condition which is more natural for Manacher arrays and their derivatives, and better captures their structure. We proceed to define the condition:
\begin{definition}[Smooth array]
    Let $\Arr$ be an array with $\ell$ integral elements from the range $[0\dd \ell-1]$. We say that the array is {\em smooth} if it satisfies that
    \[
    \forall i<j\quad \min(\Arr[i],\,\Arr[j])\le j-i.
    \]
    
\end{definition}

\begin{observation}
    Any smooth array admits a linear log-sum.
\end{observation}
This observation follows from the definition of smooth arrays, which requires values greater than $2^{m}$ to be stored at least $2^m$ indices apart. Consequently, there are at most $\ell\cdot2^{-m}$ such values, and the log-sum satisfies:
\[
\sum_{i=1}^\ell \log (\Arr[i]+1)\le \ell\sum_{m=1}^{\lceil \log \ell\rceil} m\cdot 2^{-m} =\Ohs{\ell}.
\]
\begin{corollary}~\label{cor:csmooth}
    It is possible to SDC-encode any constant number of smooth arrays, each with $\ell$ elements, using $\Oh{\ell}$ bits in total.
\end{corollary}

\section{Framework and Further Definitions}\label{s:framework}

In this section, we introduce the terminology and notation for periodic palindromes, palindromic runs, and their centers. These definitions will be used later in the construction of the compressed Manacher array.

A \emph{periodic palindrome} is a string that is both periodic and palindromic. 
Such objects play a central role in our construction. We begin by citing a useful structural property:
\begin{lemma}[Canonical decomposition of periodic palindromes~\cite{Kosolobov2015}]\label{lem:per-pal-struc}
    Any periodic palindrome $P$ admits a unique decomposition
 $(q_0q_1)^kq_0$, where:
    \begin{enumerate}
        \item $q_0$ and $q_1$ are palindromes,
        \item $\root=q_0q_1$ is the root of $P$,
        \item $P \bmod |\root|=|q_0|$,
        \item $k\geq 2$.
    \end{enumerate}
    In particular, condition (3) implies that $|q_1|\ge 1$.\\
    We call this decomposition the {\em canonical decomposition} of $P$, and $q_0q_1$ is the canonical decomposition of $\root$.
\end{lemma}

Although well defined, periodic palindromes pose challenges for our objective. For instance, consider the string $\mathttt{ababab}$. The string has the two following maximal periodic subpalindromes: $\Pal_3=[1\dd 5]$ and $\Pal_4=[2\dd 6]$. The subpalindromes overlap and have different corresponding roots. To simplify our framework, we make use of a different, more robust notion: \emph{palindromic run}, which is the periodic extension of a periodic palindrome.
\begin{definition}[Palindromic Run]\label{def:pal-rep}
    Let $S$ be a string, and let $\Run$ be a run of $S$ with root $\root$. The run $\Run$ is said to be a {\em palindromic run} if $\root=q_0q_1$ for some palindromes $q_0$ and $q_1$.
\end{definition}
Lemma 1 in~\cite{Kosolobov2015} guarantees that the decomposition $\root=q_0q_1$ with $|q_1|>0$ is unique when it exists. We say that two palindromic runs are the same (denoted by $\Run = \Run'$) if their underlying intervals are equal.

In this work, we encode palindromic runs as {\em palindromic run descriptor} ($\PPD$).
\begin{definition}[Palindromic run descriptor ($\PPD$)] \label{def:ppd}
    Let \( \Run=[i\dd j] \) be a palindromic run of $S$ with root $\root=q_0q_1$. The \emph{palindromic run descriptor} ($\PPD$) of \( \Run \) is the tuple $\PPDc,$
    where:
    \begin{itemize}
        \item \( i \) is the starting index of the interval $\Run$.
        \item \( q_0,\,q_1 \) are the palindromic factors from the decomposition of $\root$~(\cref{lem:per-pal-struc}).
        \item \( k,\,e \) are the unique integers satisfying $S[i\dd j]\;=\;\root^k\root[1\dd e]$, and $0\le e<|\root|$.
    \end{itemize}
\end{definition}
The value $\exp(\Run) \coloneq k+\frac{e}{|\root|}$ is commonly referred to as the \emph{exponent} of $\Run$.

Next, we categorize the centers in a given string. Recall from the preliminaries that the maximal subpalindrome centered at $c$ is $\Pal_c$, the palindromic extension of $\Pal_c$ (when exists) is $\Run_c$, and the root of $\Run_c$ is $\root_c$. First, we define a {\em periodic center} as a center whose associated maximal subpalindrome $\Pal_c$ is a repetition.

Among all centers inside a palindromic run, those aligned with occurrences of the palindromic factors of the root (i.e., $q_0,\,q_1$) are particularly interesting as they satisfy several useful combinatorial properties. Let $\Run=[i\dd j]$ be a palindromic run with root $\root=q_0q_1$. If a center $c$ coincides with the center of an occurrence of $q_0$ or $q_1$, we say that $c$ is a {\em repeat center of $\Run$}. 
Set $c_1 \coloneq i + \frac{|q_0|-1}{2}$ to be the first repeat center of $\Run$. Formally, the set of repeat centers of $\Run$ is
\[
C_{\mathrm{rep}}(\Run)
 = \bigl\{ c_1 + t \cdot \frac{|\root|}{2}
      \;\bigm|\; t\in\mathbb{Z}_{\ge 0} ,\;
                 i-\frac12 \le c_1 + t \cdot \frac{|\root|}{2}
                               \le j+\frac12
    \bigr\}.
\]
More generally, we say that a center $c$ is an {\em internal center of $\Run$} if it coincides with the center of an occurrence of $q_0$ or $q_1$ {\em other than} the first or last occurrences of each. The number of internal centers of $\Run$ is exactly $|C_{\mathrm{rep}}(\Run)|-4$, and since for every palindromic run $|C_{\mathrm{rep}}(\Run)|\ge 5$, at least one internal center exists. Notice that for every internal center $c$ of $\Run$, the length of $\Pal_c$ satisfies $|\Pal_c|\ge 2|\root|$.
Lastly, we observe that if $c$ is a periodic center, then it is an internal center of $\Run_c$.

\begin{example}
    Consider the string $S=\mathttt{baaab}$, with the palindromic run $\Run=[2\dd 4]$. The repeat centers of $\Run$ are $\{1.5,\,2,\,\dots,\,4,\,4.5\}$, for a total of seven centers. Of them, $\{2.5,\,3,\,3.5\}$ are the internal centers of $\Run$. Although $c=3$ is an internal center of $\Run$, it is not a periodic center as the maximal subpalindrome $\Pal_3=[1\dd 5]$ is not a repetition. Therefore, the periodic centers of $S$ are $\{2.5,\,3.5\}$.
\end{example}

The last definition in this section is the center-period array, an array that stores for every periodic center its associated period.
\begin{definition}[Center-period array]\label{def:per-len-arr}
    Let $S$ be a string.  The \emph{center-period array} $\mathsf{L}$ assigns to (i) each periodic center $c$ the period of its associated palindromic run, i.e. $|\root_c|$, and (ii) zero to all other centers.
    Formally, the center-period array $\mathsf L$ is defined as:
    \[
    \mathsf{L}[2c] =
        \begin{cases}
        0 & \text{if $c$ is not a periodic center}, \\
        |\root_c| & \text{otherwise}.
        \end{cases}
    \]
\end{definition}

\newcommand{\qz}{q_0^{(4)}}
\newcommand{\qo}{q_1^{(4)}}
The following example summarizes all the definitions of this section.
\begin{example}\label{ex:all-def}
    We demonstrate the main definitions in this section: palindromic run, palindromic run descriptor, periodic centers and internal centers (of a run), and the center-period array. Let $S$ be a string of length $29$:
    \[
    S=\mathttt{xxabababacddcababacddcabababy}.
    \]

    There are five palindromic runs in $S$:
    \[
    \Run=[5\dd 27],\;\Runi{1}=[3\dd 9],\; \Runi{2}=[14\dd 18],\; \Runi{3}=[23\dd 28],\; \Runi4=[1\dd 2].
    \]
    The runs $\Run,\,\Runi2$ and $\Runi4$ all have five repeat centers, hence one internal center. These centers are $16,\,16$ and $1.5$, respectively. The internal centers of $\Runi1$ are $\{5,\,6,\,7\}$ and the internal centers of $\Runi3$ are $\{25,26\}$. The runs and their respective roots, repeat centers and internal centers are drawn in~\cref{subfig:pal-run}.
    \begin{figure}[h!]
        \centering
        \begin{tikzpicture}[x=0.35cm,y=0.8cm,>=stealth]

    \begin{scope}[shift={(12,5)}]

      \draw[rounded corners] (0,0.4) rectangle (15,-1.3);
      \node[anchor=west] at (0.2,0.1) {\scriptsize Legend};

      \draw[fill=black,draw=none] (0.4,-0.4) circle (1.5pt);
      \node[anchor=west] at (0.8,-0.4) {\scriptsize repeat center};
    
      \draw[red] (0.4,-0.9) circle (3pt);
      \node[anchor=west] at (0.8,-0.9) {\scriptsize internal center};
    
      \draw[blue] (7.4,-0.3) rectangle (7.8,-0.5);
      \node[anchor=west] at (7.8,-0.4) {\scriptsize occurrence of $q_0$};
    
      \draw[black!50!green] (7.4,-0.8) rectangle (7.8,-1);
      \node[anchor=west] at (7.8,-0.9) {\scriptsize occurrence of $q_1$};
    \end{scope}

    \node[above] at (12, 2.75) {The palindromic run $\Run=[5\dd 27]$. $q_0=\mathttt{ababa},\quad q_1=\mathttt{cddc}$.};
    
    \foreach \x [count=\i from 0] in {x,x,a,b,a,b,a,b,a,c,d,d,c,a,b,a,b,a,c,d,d,c,a,b,a,b,a,b,y}
        \node[draw=none, fill=none, minimum size=7mm] (A\i) at (\i,2) {\mathttt{\x}};

    \foreach \j in {11, 20} {
        \draw[fill=black,draw=none] ({\j-0.5},2.5) circle (1.2pt);
    }
    \foreach \j in {6, 15, 24} {
        \draw[fill=black,draw=none] ({\j},1.5) circle (1.2pt);
    }
    \foreach \j in {15} {
        \draw[red] (\j,1.5) circle (3pt);
    }

    \foreach \y in {8, 17}
        \draw[black!50!green] ({\y+0.5},1.75) rectangle ({\y+4.5},2.25);
    
    \foreach \y in {3, 12, 21}
        \draw[blue] ({\y+0.5},1.75) rectangle ({\y+5.5},2.25);
\end{tikzpicture}

\vspace{1em}

\begin{tikzpicture}[x=0.35cm,y=0.8cm,>=stealth]

    \node[above] at (12, 3) {For all the runs with period $2$, $q_0=\mathttt{a},\quad q_1=\mathttt{b}$.};

    \node[above] at (5, 2.25) {$\Runi1=[3\dd9]$};
    \node[above] at (15.5, 2.25) {$\Runi2=[14\dd18]$};
    \node[above] at (24.5, 2.25) {$\Runi3=[23\dd28]$};
    \foreach \x [count=\i from 0] in {x,x,a,b,a,b,a,b,a,c,d,d,c,a,b,a,b,a,c,d,d,c,a,b,a,b,a,b,y}
        \node[draw=none, fill=none, minimum size=7mm] (A\i) at (\i,2) {\mathttt{\x}};
    
    \foreach \j in {1, 2, 3, 4, 5, 6, 7, 12, 13, 14, 15, 16, 21, 22, 23, 24, 25, 26} {
        \draw[fill=black,draw=none] ({\j+1},1.5) circle (1.2pt);
    }

    \foreach \j in {3, 4, 5, 14, 23, 24} {
        \draw[red] ({\j+1},1.5) circle (3pt);
    }

    \foreach \y in {1,3,5,7,12,14,16,21,23,25}
        \draw[blue] ({\y+0.5},1.75) rectangle ({\y+1.5},2.25);
    
    \foreach \y in {2, 4, 6, 13, 15, 22, 24, 26}
        \draw[black!50!green] ({\y+0.5},1.75) rectangle ({\y+1.5},2.25);
\end{tikzpicture}

\vspace{1em}

\begin{tikzpicture}[x=0.35cm,y=0.8cm,>=stealth]
    \node[above] at (12, 2.75) {The palindromic run $\Runi4=[1\dd 2]$. $q_0=\emptystring,\quad q_1=\mathttt{x}$.};
    \foreach \x [count=\i from 0] in {x,x,a,b,a,b,a,b,a,c,d,d,c,a,b,a,b,a,c,d,d,c,a,b,a,b,a,b,y}
        \node[draw=none, fill=none, minimum size=7mm] (A\i) at (\i,2) {\mathttt{\x}};

    \foreach \j in {0, 1, 2} {
        \draw[fill=black,draw=none] ({\j-0.5},2.5) circle (1.2pt);
    }
    
    \foreach \j in {1} {
        \draw[red] ({\j-0.5},2.5) circle (3pt);
    }
    \foreach \j in {0, 1} {
        \draw[fill=black,draw=none] (\j,1.5) circle (1.2pt);
    }

    \foreach \y in {0, 1}
        \draw[black!50!green] ({\y-0.5},1.75) rectangle ({\y+0.5},2.25);

\end{tikzpicture}
        \caption{Visual demonstration of repeat centers and internal centers. At the top, the palindromic run with period nine is described, below the three runs with period two, and lastly the run with period one.}
        \label{subfig:pal-run}
    \end{figure}

    Although $16$ is an internal center of both $\Runi2$ and $\Run$, it is not a periodic center, as $\Pal_{16}=[4\dd 28]$ is not a repetition. The other internal centers are all periodic. An illustration of the center-period array of $S$ together with the Manacher array can be found in~\cref{fig:cent-per}.

    \begin{figure}[h!]
        \centering
        \begin{tikzpicture}[xscale=0.38]

\def\labelX{-2.25}
\def\Aeb{0.9}
\def\Leb{0.4}

\foreach \x [count=\i from 0] in {x,x,a,b,a,b,a,b,a,c,d,d,c,a,b,a,b,a,c,d,d,c,a,b,a,b,a,b,y}
    \node[draw=none, fill=none, minimum size=7mm] (A\i) at (\i,2) {\mathttt{\x}};

\foreach \y in {1,5, ..., 29}{
  \node[black!70, draw=none, fill=none, minimum size=7mm] at ({\y-1},1.5) {\scriptsize \y};
}

\foreach \j/\y in {
  0/0, 1/0, 2/0, 8/0, 9/0, 10/0, 11/0, 12/0, 13/0,
  17/0, 18/0, 19/0, 20/0, 21/0, 22/0, 27/0, 28/0
}{
  \node[minimum size=7mm] at (\j,\Aeb) {\textcolor{black!10}{0}};
}

\foreach \j/\y in { 3/1, 7/1, 4/2, 5/3, 6/2, 14/1, 15/12, 16/1, 23/1, 24/2, 25/2, 26/1} {
  \node[minimum size=7mm] at (\j,\Aeb) {\y};
}

\foreach \j in {
  -1,1,2,3,4,5,6,7,8,9,
  11,12,13,14,15,16,17,18,
  20,21,22,23,24,25,26,27,28
}{
  \node[minimum size=7mm] at ({\j+0.5},2.5) {\textcolor{black!10}{0}};
}

\foreach \j/\y in { 0/1, 10/7, 19/7 }{
  \node[minimum size=7mm] at ({\j+0.5},2.5) {\y};
}

\foreach \i in {
  0,1,2,3,7,8,9,10,11,12,13,14,15,16,17,18,19,20,21,22,23,26,27,28
}{
  \node[minimum size=7mm] at (\i,\Leb) {\textcolor{black!10}{0}};
}

\foreach \i/\v in {4/2, 5/2, 6/2, 24/2, 25/2}{
  \node[minimum size=7mm] at (\i,\Leb) {\v};
}

\foreach \y [count=\j from 2] in {0, 0, 0, 0, 0, 0, 0, 0, 0, 0, 0, 0, 0, 0, 0, 0, 0, 0, 0, 0, 0, 0, 0, 0, 0, 0, 0, 0} {
    \node[black!10, draw=none, fill=none, minimum size=7mm] (C\j) at ({\j-0.5},3) {\y};
}
\node[black!10, draw=none, fill=none, minimum size=7mm] at ({-0.5},3) {0};
\node[draw=none, fill=none, minimum size=7mm] at ({0.5}, 3) {1};

\draw[blue] ({2.5},1.75) rectangle ({27.5},2.25);

\foreach \y in {3, 21}
\draw[red] ({\y+0.5},1.8) rectangle ({\y+2.5},2.2);

\foreach \y in {0}
\draw[red] ({\y-0.5},1.8) rectangle ({\y+0.5},2.2);

\node[draw=none, fill=none, minimum size=7mm] (Label) at (\labelX,\Leb) {$\mathsf L_{\text{even}}$};

\node[draw=none, fill=none, minimum size=7mm] (Label) at (\labelX,\Aeb) {$\mathsf A_{\text{even}}$};

\node[draw=none, fill=none, minimum size=7mm] (Label) at (\labelX,2.5) {$\mathsf A_{\text{odd}}$};

\node[draw=none, fill=none, minimum size=7mm] (Label) at (\labelX,3) {$\mathsf L_{\text{odd}}$};

\end{tikzpicture}
        \caption{A visualization of the center-period array $\mathsf L$ and the Manacher array $\mathsf A$. For each periodic center, the root $\root_c$ is highlighted in red. The maximal palindrome is highlighted in blue. For clarity, each array is split into two subarrays -- the values in odd and in even indices. }
        \label{fig:cent-per}
    \end{figure}
    
    We conclude by showing the $\PPD$ of the palindromic runs:
    \begin{align*}
        \PPD=\PPDc \quad &\PPD(\Run)=(5,5,4,2,5) \quad \PPD(\Runi{4})=(1,0,1,2,0)\\
        \PPD(\Runi{1})=(3,1,1,3,1)\quad 
        & \PPD(\Runi{2})=(14,1,1,2,1) \quad \PPD(\Runi{3})=(23,1,1,3,0) \\
    \end{align*}
\end{example}

\section{Properties of Periodic Palindromes}

Using the terminology introduced in the previous section, we develop the structural properties of palindromic runs and periodic palindromes. \\

Our starting point is the following lemma:
\begin{lemma}[Lemma 3 in~\cite{Kosolobov2015}]\label{lem:cent-overlap}
    Let $P$ be a periodic palindrome and suppose $(q_0q_1)^kq_0$ is the canonical decomposition of $P$.
    If $\mathcal U=[i\dd j]$ is a subpalindrome of $P$ such that $|\mathcal U| \ge |q_0q_1|-1$, then the center of $\mathcal U$ coincides with the center of some $q_0$ or $q_1$ from the decomposition.
\end{lemma}

Next, we prove a vital structural property: a maximal periodic subpalindrome (i.e., a maximal subpalindrome that is periodic) is a prefix or a suffix of its periodic extension.
\begin{lemma}\label{lem:no-ext-pal}
    Let $S$ be a string, $\Pal=[i\dd j]$ be a maximal periodic subpalindrome, and $\Run=[i'\dd j']$ be the periodic extension of $\Pal$. Then $i=i'\;\text{ or }\;j=j'$.
\end{lemma}
\begin{proof}\label{ap:pr:lem:no-ext-pal}
    Let $\root = q_0q_1$ be the root of $\Pal$.  
    By~\cref{lem:per-pal-struc}, $S[i\dd j] = (q_0q_1)^k q_0$ for some $k$, hence $j+1=i+k|\root|+|q_0|$.
    Since $\Run$ is the periodic extension of $\Pal$, the run $\Run$ has the same period $|\root|$, and the interval borders satisfy $i\ge i'$ and $j \le j'$.
    
    Assume for contradiction that $i > i'$ \textbf{and} $j < j'$.  
    In this case, $\Pal$ is strictly contained in $\Run$, and in particular $i > 1$ and $j < |S|$.
    
    Because $\Pal$ is a maximal subpalindrome, its endpoints cannot be extended:  
    \begin{equation}\label{eq:noexta}
    S[i-1] \neq S[j+1].
    \end{equation}
    
    On the other hand, since $\Run$ has period $|\root|$, we are guaranteed that $S[t] = S[t+|\root|]$ for all $i' \le t$ and $t+|\root| \le j'$. 
    By plugging $t=i-1$, we get
    \[
    S[i-1] = S[i-1+|\root|] = \root[|\root|]=q_1[|q_1|].
    \]
    Applying the periodicity property $k$ times to $t=i+|q_0|$, we get
    \[
    S[j+1] = S[i+k|\root|+|q_0|]=S[i+|q_0|] = q_1[1].
    \]
    
    Since $q_1$ is a palindrome, $q_1[1]= q_1[|q_1|]$, hence $S[i-1] = S[j+1]$, contradicting~\eqref{eq:noexta}, and our assumption was false.
\end{proof}

A corollary of this lemma is that given a periodic center $c$, the radius of $\Pal_c$ is the distance to the closer border of $\Run_c$.

We prove another key structural fact, showing that if a palindrome has a sufficiently long periodic factor, then the palindrome itself is periodic.
\begin{lemma}\label{lem:no-ext-per}
    Let $P$ be a palindrome, and $\mathcal T=[2\dd |P|]$ be a repetition with root $\root$. The period of $P$ is $|\root|$.
\end{lemma}
\begin{proof}\label{ap:pr:no-ext-per}
    To prove the lemma, we show that $P[1]=\root[|\root|]$, hence $\mathcal T$ can be extended without changing its period. We consider two cases: when $|\mathcal T|=2|\root|$, and when $|\mathcal T|\ge 2|\root|+1$.

    In the first case, $P[2\dd |P|]=\root\root$. Since $P$ is a palindrome, $P[1]=P[|P|]=\root[|\root|]$.
    
    In the second case, the length of the subpalindrome $[2\dd |P|-1]$ is at least $2|\root|$, hence it's a periodic subpalindrome. According to~\cref{lem:per-pal-struc}, the root $\root$ can be written as $\root=q_0q_1$ and the factor $P[2\dd |P|-1]$ can be written as $(q_0q_1)^kq_0$. Furthermore:
    \begin{alignat*}{2}
        & P[|P|]=q_1[1] && \qquad \text{($\mathcal T$ is a repetition with period $|q_0q_1|$)}\\
        & P[|P|]=P[1] && \qquad \text{($P$ is a palindrome)}\\
        & q_1[1]=q_1[|q_1|] && \qquad \text{($q_1$ is a palindrome)}
    \end{alignat*}
    Using the above equalities, $P[1]=q_1[|q_1|]=\root[|\root|]$, as required.
\end{proof}

We conclude the section by proving that every internal center $c$ of a palindromic run $\Run$ yields $\Run_c = \Run$, with the sole exception of $c=c_\Run$.
\begin{lemma}\label{lem:only-center}
    Let $S$ be a string, $\Run$ be a palindromic run, and $c$ be an internal center of $\Run$. At least one of the following is satisfied:
    \begin{enumerate}[i]
        \item The maximal subpalindrome $\Pal_c$ is a prefix or a suffix of $\Run$, or
        \item $\Pal_c$ and $\Run$ share the same center (i.e., $c_{\Pal_c}=c_\Run$).
    \end{enumerate}
\end{lemma}
\begin{proof}[Proof for~\cref{lem:only-center}]\label{ap:pr:lem:only-center}
    The claim follows from~\cref{lem:no-ext-per}. Let $\root=q_0q_1$ be the root of $\Run$.

    First, notice that the maximal subpalindrome centered at an internal center extends at least to the boundary of $\Run$, and its length when extended to the border of $\Run$ is at least $2|\root|+|q_0|\ge2|\root|$. To violate the lemma's conditions, $\Pal_c$ cannot border with $\Run$, hence $\Pal_c$ is not a sub-interval of $\Run$ and its length is at least $2|\root|+1$. Denote $c_P\coloneq c_{\Pal_c}$, and assume w.l.o.g that $c_P<c_\Run$.

    Let $i$ (resp. $i'$) be the starting index of $\Pal_c$ (resp. $\Run$), let $j'$ be the ending index of $\Run$, and set $t = c_P - i' + 1$.  
    Because $c_P < c_\Run$, we have $i < i'$.  
    
    Observe that the factor $S[c_P - t \dd c_P + t]$ is palindromic, as its radius is no larger than that of $\Pal_c$. 

    For its right endpoint, we compute:
    \[
    c_P + t = c_P + (c_P - i' + 1) = 2c_P - i' + 1 \le 2c_\Run - i' = 2\frac{j'+i'}{2} - i' = j',
    \]
    and for its left endpoint:
    \[
    c_P - t = c_P - (c_P - i' + 1) = i' - 1.
    \]

    Therefore, set:
    \[
        \Pal:=[c_P-t\dd c_P+t]\quad\text{(subpalindrome)},\quad \mathcal T:=[c_P-t+1\dd c_P+t]\quad\text{(has period $|\root|$)}.
    \]
    The subpalindrome $[c_P-t+1\dd c_P+t-1]$ is a prefix of $\Run$. Since $c_P$ is an internal center of $\Run$, the length of that subpalindrome is at least $2\len{\root}$, hence $\len{T}>2\len{\root}$, and~\cref{lem:no-ext-per} implies that $\Run$ is not a maximal repetition, hence it is not a palindromic run, contradicting the lemma's assumptions.
\end{proof}

\begin{corollary}\label{cor:at-most-one}
    Let $c_1,\,c_2$ be two internal centers of a palindromic run $\Run$. At least one of the centers is periodic, and its palindromic run is $\Run$, i.e., $\Run_c=\Run$.
\end{corollary}

\section{Compact Manacher Array}\label{s:fast-access}

\newcommand{\CP}{\mathsf L}
\newcommand{\TArr}{\tilde{\Arr}}
This section introduces a compressed data structure for the Manacher array, ultimately proving~\cref{th:main}. Throughout the section, we denote by $S$ the input string of length $n$, by $\Arr$ its Manacher array, and by $\CP$ its center-period array.
\begin{remark} For clarity, we assume throughout this section that $S$ has no even-length subpalindromes. This simplifying assumption {\em does not affect the generality of the result}, but ensures consistent and clearer notation. Specifically, a subpalindrome $\Pal$ of $S$ with radius $\mathrm{rad}(\Pal)$ and center $c_\Pal$ corresponds to the factor $S[c_\Pal-\mathrm{rad}(S)\dd c_\Pal+\mathrm{rad}(S)]$, and $c_\Pal$ is guaranteed to be an integer.
\end{remark}

\subsection{High-level idea}
The idea underlying the data structure is to store information about periodic and non-periodic centers in separate, smooth arrays, and encode those arrays using SDC, supporting the required time and space bounds.
\begin{enumerate}
    \item Initially, we show that setting the radii at periodic centers in $\Arr$ to zero guarantees that the resulting array $\TArr$ is smooth~(\cref{lem:main-array}).
    \item We proceed by showing that the center-period array $\CP$ is smooth~(\cref{lem:per-len}).
    \item We then describe a data structure $\I$, that given an arbitrary periodic center $c$ locates the first internal center of $\Run_c$ in constant time~(\cref{lem:layb}).
    \item Lastly, we show that all of these structures suffice to compute the radius of the maximal subpalindrome at any given center in constant time~(\cref{lem:radius-calc}).
\end{enumerate}

\subsection{Description of the structures}
The first data structure employed is the {\em sparse Manacher array}.
\begin{definition}[Sparse Manacher array]\label{def:sparse-man}
    Given a string $S$ and its Manacher array $\Arr$, the {\em sparse Manacher array} $\TArr$ is defined by:
    \[
      \TArr[2c] =
      \begin{cases}
        0 & \text{if $c$ is a periodic center,}\\
        \Arr[2c]             & \text{otherwise}.
      \end{cases}
    \]
\end{definition}

\begin{lemma}\label{lem:main-array}
    The sparse Manacher array of $S$ is smooth.
\end{lemma}

\begin{proof}\label{pr:main-array}
We prove the claim by contradiction. Let $ c_1<c_2 $ be two centers satisfying:
\[
\TArr[2c_1] \le \TArr[2c_2] \quad \text{and} \quad \TArr[2c_1] > (2c_2-2c_1).
\]
Denote $ d \coloneq c_2-c_1 $. Substituting $d$ into the expression, the condition $ \TArr[2c_1] > 2d $ implies a significant overlap between the maximal subpalindromes at centers $c_1,\,c_2$, i.e., $\Palc1,\,\Palc2$. More precisely:
\begin{enumerate}[i]
    \item $[c_1-2d\dd c_1+2d]$ is a subpalindrome centered at $c_1$, and
    \item $[c_2-d\dd c_2+d]$ is its palindromic suffix centered at $c_2$. 
\end{enumerate}
Consequently~(\cref{p:pal-per}), this overlap implies that the factor $S[c_1-2d\dd c_1+2d]$ is periodic with period $(4d+1)-(2d+1)=2d$.

By invoking~\cref{lem:cent-overlap}, both centers $c_1,\,c_2$ are internal centers of some palindromic run. However, by~\cref{cor:at-most-one}, at least one of them is a periodic center and would have been set to zero, contradicting the definition of $\TArr$.
\end{proof}

Next, we show how to compute values from the Manacher array at periodic centers. We begin by claiming that the center-period array is smooth.

\begin{lemma}\label{lem:per-len}
    The center-period array $\CP$ of a string $S$ is smooth.
\end{lemma}
\begin{proof}[Sketch] 
    The high-level idea supporting the proof is that internal centers of different palindromic runs that occur ``too close'' to one another force a common period for both runs. \\
    The proof proceeds by contradiction. We assume two centers $c_1,\,c_2$ violate the smoothness condition. We analyze their palindromic runs $\Runc1$ and $\Runc2$, dividing the problem into three cases based on alignment and containment. \\
    \textbf{Case 1: $\Runc1 = \Runc2$}: Periodic centers of the same run inherently satisfy the smoothness condition. \\
    \textbf{Case 2 (Containment)}: If $c_1$ is an internal center of $\Runc2$, it forces a synchronization of periods, and if $c_1$ is not an internal center of $\Runc2$ then its period is sufficiently small and cannot violate the smoothness condition with $c_2$. \\
    \textbf{Case 3 (Partial or no overlap).}
    We assume that $\len{\rootc1}\le \len{\rootc2}$ and $c_1<c_2$. Consider $c_1,c_2$ and their adjacent repeat centers along the runs,
    \[
      c_1^{(+1)},\,c_1^{(+2)},\,c_2^{(-1)},\,c_2^{(-2)}.
    \]
    Using the fact that $c_1,c_2$ form a counterexample and that the runs are not nested, it follows that all four of these centers lie inside both $\Run_1$ and $\Run_2$. In particular, there is a subpalindrome of length $\len{\rootc1}$ centered at $c_2^{(-1)}$, which together with~\cref{lem:cent-overlap} yields a subpalindrome of length $\len{\rootc2}$ centered at $c'\coloneq c_2^{(-1)}+\tfrac{1}{2}\len{\rootc1}$. Thus $c_2^{(-1)}$ and $c'$ are repeat centers shared by $\Runc1$ and $\Runc2$; in particular, both runs have the same period, and the runs synchronize, which was covered by Case~1. The above argument also rules out the disjoint configuration. 
\end{proof}
The formal proof, together with a visual demonstration of the different cases, is deferred to~\cref{ss:in-depth} as~\cref{pr:lem:per-len}.

\newcommand{\TI}{\tilde{\I}}
\newcommand{\TL}{\tilde{\CP}}
So far, we have shown that the arrays storing (i) radii at non-periodic centers and (ii) periods at periodic centers are both smooth, and can be encoded using SDC, supporting constant-time access. The missing piece for the complete data structure is to compute the radius at a periodic center. The primary obstacle to doing so is locating the borders of the palindromic run of the given center. The following lemma resolves this structural challenge by introducing an auxiliary data structure $\I$ that identifies the first internal center of a run given any internal center of the same run. This constitutes the final non-trivial construction; the derivation of the radius from this center is mechanical and is detailed subsequently.

\begin{lemma}\label{lem:layb}
    Given a string $S$ of length $n$, there exists a data structure $\I$ occupying $\Oh{n}$ bits that supports constant-time queries for the first (lowest) internal center of $\Run_c$ for any periodic center $c$.
\end{lemma}

\begin{proof}[Sketch of construction]
    We build a hierarchy of bitvectors $\I[1],\I[2],\dots$ of exponentially
    decreasing length. The $m$-th bitvector $\I[m]$ has $\Theta(2n\cdot2^{m-1})$ bits
    and is responsible for runs whose period $p$ lies in the range
    $[2^{m-1},2^m)$. For such a run $\Run$ with first internal center $c_1$, we
    store a single `1' in $\I[m]$ at position
    \[
      \scl{c_1}{m} \;=\; \shr{2c_1}{m-1},
    \]
    and all other positions remain `0'. Each $\I[m]$ is preprocessed for
    \mathttt{rank}/\mathttt{select} queries. Since the lengths of successive layers
    shrink geometrically and the first layer has $\Oh{n}$ bits, the total size is $\Oh{n}$.
    
    Given a periodic center $c$, we read its period $p_c$ from the center–period
    array $\CP$ and choose the corresponding layer $m$ with
    $2^{m-1} \le p_c < 2^m$. We then map $c$ to its scaled position
    $c_m = \scl{c}{m}$ and perform a predecessor search among the `1'-bits of
    $\I[m]$ at position $c_m$ using a single \mathttt{rank} followed by a
    \mathttt{select}. This predecessor gives us the scaled position of the first
    internal center $c_1$, which we can recover in constant time by inverting the
    scaling and using the periodicity of $\Run_c$. The smoothness of $\CP$ ensures
    that at most one candidate center in the relevant range is compatible with the
    period $p_c$, so the recovered center is indeed the first internal center of
    $\Run_c$.
\end{proof}

The full construction, together with correctness proof, is deferred to~\cref{ss:in-depth} as~\cref{pr:lem:layb}.

We now have all the required pieces to compute the radius of the maximal subpalindrome at any given center. We formally present the final data structures: 
\begin{definition}[Enriched center-period array]\label{def:enriched-cent-per}
        The {\em Enriched Center-Period Array} $\TL$ is an extension to the center-period array. For each periodic center $c$ with $\PPD=\PPDc$, the array $\TL$ stores the triplet $(\len{\root_c},\, e,\, \len{q_0})$. 
\end{definition}
Since $e, \len{q_0} < \len{\root_c}$, these values can be stored as three different smooth arrays\footnote{Strictly speaking, storing the triplet for every periodic center is suboptimal but does not affect the asymptotic space bounds.}. \\
The final data structure is the boundary locator.
\begin{definition}[Boundary locator]\label{def:bound-loc}
    The {\em boundary locator} $\TI$ is a composite data structure consisting of the layered bitvector $\mathbf{I}$ (from \cref{lem:layb}) and a symmetric data structure {\em $\mathbf{I}_{\text{last}}$}, that given an arbitrary periodic center $c$ locates the \emph{last} (highest) internal center $c_m$ of the run $\Run_c$.
\end{definition}

The following lemma assembles the pieces and explains how to find the radius given a periodic center.
\begin{lemma}\label{lem:radius-calc}
    Let $c$ be a periodic center of $S$. The radius of the maximal subpalindrome centered at $c$ can be found in constant time, using $\Ohs{n}$ bits of space.
\end{lemma}

\begin{proof}
    To compute the radius, we first reconstruct the $\PPD$ of the palindromic run $\Run_c$, using $\TL$ and $\TI$.

    \noindent\textbf{Algorithm.} The retrieval proceeds in three steps:
    \begin{enumerate}
        \item \textbf{Find Boundaries:} Query $\TI$ with $c$ to obtain $c_1$ and $c_m$.
        \item \textbf{Retrieve Parameters:} Query $\TL[2c]$ to retrieve the period $\len{\root_c}$, the excess $e$, and $\len{q_0}$. From these, we compute the suffix length $\len{q_1} = \len{\root_c} - \len{q_0}$.
        \item \textbf{Compute $\PPD$:} Using the retrieved boundaries and period length, we compute the number of repetitions $k$:
        \[
             k = 1+\left\lceil \frac{c_m - c_1}{\len{\root_c}}\right\rceil+\begin{cases}
                 1 & \text{if } e<\len{\root_c}-\frac{\len{q_1}}{2}, \\
                 0 & \text{otherwise.}
             \end{cases}
        \]
        The computation of $k$ diverges by values of $e$, as the suffix $\root[1\dd e]$ introduces two repeat centers to the run $\Run$ when $e\ge |\root_c|-\frac{1}{2}\len{q_1}$.

        The starting index $i$ of the run is computed relative to the first internal center:
        \[
            i = c_1 - \len{\root_c} - \frac{\len{q_0}-1}{2}.
        \]
    \end{enumerate}
    
    \noindent\textbf{Radius Calculation.} With the tuple $\PPDc$ fully determined, the boundaries of the palindromic run $\Run_c$ are known. Thus, the radius is simply the distance from $c$ to the nearest run boundary. (Recall~\cref{lem:no-ext-pal}). 

    Since both $\TI$ and $\TL$ consume $\Ohs{n}$ bits and support constant time query, and all arithmetic operations are constant time, the lemma holds.
\end{proof}

\subsection{The complete data structure}
We can now combine the components described in the previous subsections to prove our main result. We restate the main theorem:
\begingroup

  \renewcommand{\thetheorem}{\ref{th:main}}
    \begin{theorem}
        Let $S$ be a string of length $n$, and let $\mathsf{A}$ denote its Manacher array. The array $\mathsf{A}$ admits a lossless encoding of size $\Ohs{n}$ bits that supports element access in constant time.
    \end{theorem}
  \addtocounter{theorem}{-1}
\endgroup

\begin{proof}
    The data structure consists of three components:
    \begin{enumerate}
        \item $\TArr$: The sparse Manacher array, storing radii for all non-periodic centers, and zero for periodic centers (\cref{def:sparse-man}).
        \item $\TL$: The enriched center-period array, storing run details ($\len{\root},\,\len{q_0},\,e)$ for periodic centers (\cref{def:enriched-cent-per}).
        \item $\TI$: The boundary locator, computing the first/last internal center of a palindromic run given one of its internal centers (\cref{def:bound-loc}).
    \end{enumerate}
    
    As established in the respective lemmas, each component occupies $\Ohs{n}$ bits. Thus, the total space usage is $\Ohs{n}$ bits.

    To retrieve the radius at an arbitrary center $c$, the query proceeds as follows:
    First, we check $\TArr[c]$. If the value is non-zero, we return it directly.
    Otherwise, $c$ is a periodic center. We query $\TL$ to retrieve the triplet $(\len{\root_c}, \len{q_0}, e)$, and query $\TI$ to locate the run boundaries $c_1$ and $c_m$. By~\cref{lem:radius-calc}, these parameters allow us to compute the radius of the maximal subpalindrome at $c$ in constant time.
\end{proof}

\subsection{Technical details}\label{ss:in-depth}
This subsection provides the full details of the sketched proofs and constructions above.

\begin{proof}[Proof for~\cref{lem:per-len}]\label{pr:lem:per-len}
    Recall that the smoothness condition requires that:
    \[
    \forall c_1,\,c_2\quad \min(\mathsf{L}[2c_1],\,\mathsf{L}[2c_2])\le 2\abs{c_2-c_1}.
    \]

    We denote by $c_1,\,c_2$ two centers that do not satisfy the smoothness condition.

    Since $c_1,\,c_2$ violate the smoothness condition, the following is satisfied:
    \begin{equation}
        \min(\len{\rootc1},\,\len{\rootc2}) > 2\abs{c_2-c_1}.\tag{A}\label{eq:c1c2bad}
    \end{equation}
    
    We split into three cases, based on how $\Runc1$ is aligned compared to $\Runc2$. Particularly, we say that $\Runc1$ and $\Runc2$ {\em nest} if $\Runc1$ is contained in $\Runc2$ or the other way around.

    \textbf{Case 1: $\Runc1= \Runc2$.}  Denote $\rootc1=\rootc2=\root=q_0q_1$ as the common periodic factor. The repeat centers of $\Run$ coincide with occurrences of $q_0$ and $q_1$. Hence, the difference between two consecutive centers is $\frac{|q_0|+|q_1|}{2}=\frac{|\root|}{2}$.
    
    Consequently, $\min(\mathsf{L}[2c_2],\mathsf{L}[2c_1]) = |\root|$, but $2|c_2-c_1|\ge |\root|$, which follows the smoothness condition.

    \textbf{Case 2: $\Runc1$ and $\Runc2$ nest.} We assume without loss of generality that $\Runc1$ is nested inside $\Runc2$, i.e., $\Runc1\subset \Runc2$. Since $\Palc1$ is a subpalindrome, it is either shorter than $|\rootc2|-1$, or $c_1$ is a periodic center of $\Runc2$~(\cref{lem:cent-overlap}).
    
    \textit{Case 2.1: $|\Palc1|\ge |\rootc2|-1$.} In this case, $c_1,\,c_2$ are both centers of $\Runc2$, and $\Runc1= \Runc2$ which is covered by Case 1.

    \textit{Case 2.2: $|\Palc1|< |\rootc2|-1$.} In this case, $2|\rootc1|<|\rootc2|$ and $c_1$ is not an internal center of $\Runc2$.
    
    Because $c_1,\,c_2$ violate the smoothness condition, the following inequality holds:
    \[
    \abs{c_2-c_1} < \frac{1}{2}\len{\rootc1}=\frac{1}{2}\min(\len{\rootc1},\,\len{\rootc2}).\qquad \tag{i}
    \]

    Moreover, as $c_1$ is an internal center, the centers $\{c_1-|\rootc1|,\,c_1+|\rootc1|\}$ are repeat centers of $\Runc1$. Combining with (i), one of the following inequalities is satisfied:
    \[
    c_1-\frac{1}{2}\len{\rootc1} < c_2 < c_1,\quad \text{or}\quad c_1 < c_2 < c_1+\frac{1}{2}\len{\rootc1}. \tag{ii}
    \]
    Using (ii), there exists a subpalindrome centered at $c_2$ of length $\ge \len{\rootc1}$, and that subpalindrome is contained in $\Palc1$. By~\cref{lem:cent-overlap}, the center $c_2$ must be an internal center of $\Runc1$, which reduces to Case (1), and the smoothness condition is satisfied.

\newcommand{\co}[1]{c_{1}^{(+#1)}}
\newcommand{\ct}[1]{c_{2}^{(-#1)}}
    \textbf{Case 3: $\Runc1$ and $\Runc2$ do not nest.} We assume without loss of generality that $c_1<c_2$ and that $\len{\rootc1}\le \len{\rootc2}$.

    With our assumptions and notation, the violation condition~\eqref{eq:c1c2bad} can be rewritten as:
    \[
    c_2-c_1<\frac{1}{2}\len{\rootc1}\leq \frac{1}{2}\len{\rootc2}. \tag{i}
    \]

    Since all non-zero values in $\mathsf L$ come from internal centers, the following centers are repeat centers of $\Runc1$:
    \[
    \co1\coloneq c_1+\frac{1}{2}|\rootc1|,\quad \co2\coloneq c_1+|\rootc1|,
    \]
    and respectively for $\Runc2$:
    \[
    \ct2\coloneq c_2-|\rootc2|,\quad \ct1\coloneq c_2-\frac{1}{2}|\rootc2|.
    \]

    And by condition (i):
    \[
        \ct2<\ct1<c_1<c_2<\co1<\co2.
    \]

    Since $\Runc1$ and $\Runc2$ are not nested, $\ct2$ is within the borders of $\Runc1$ and $\co2$ is within the borders of $\Runc1$.

    Consider the palindrome $S[\ct2\dd c_2]$: it is contained in $\Runc1$, its length is $|\rootc2|+1$ and it is centered at $\ct1$. Therefore, by~\cref{lem:cent-overlap}, $\ct1$ is a repeat center of $\Runc1$, resulting in:
    \[
        c'\coloneq \ct1+\frac{1}{2}\len{\rootc1}\le c_1. \tag{ii}
    \]

    Now, consider the subpalindrome of length $\len{\rootc2}$ centered at $c'$. The subpalindrome starts at index
    \[
    c'-\frac{1}{2}\len{\rootc2}> \ct1-\frac{1}{2}\len{\rootc2}=\ct2,
    \]
    which is inside the boundaries of $\Runc2$, and the subpalindrome ends at index
    \[
        c'+\frac{1}{2}\len{\rootc2}=\ct1+\frac{1}{2}\len{\rootc1}+\frac{1}{2}\len{\rootc2}=c_2+\frac{1}{2}\len{\rootc1}<\co1+\frac{1}{2}\len{\rootc1}=\co2,
    \]
    which is also inside the boundaries of $\Runc2$. Therefore, by invoking~\cref{lem:cent-overlap} again, both $c'$ and $\ct1$ are repeat centers of $\Runc2$. Therefore:
    \[
    \frac{1}{2}\len{\rootc2}\le c'-\ct1 =\frac{1}{2}\len{\rootc1}\le\frac{1}{2}\len{\rootc2},
    \]
    implying that $\len{\rootc1}=\len{\rootc2}$. However, the size of the shared interval $\Runc1 \cap \Runc2$ is greater than $\len{\rootc1}$, and the palindromic runs synchronize, resulting in $\Runc1=\Runc2$.
\end{proof}

\begin{figure}
    \centering
\newcommand{\rheight}{0.5}
\newcommand{\Alength}{14}
\newcommand{\qzlength}{1.0}
\newcommand{\qolength}{1.6}
\pgfmathsetmacro{\tlength}{\qolength+\qzlength}
\begin{subfigure}{\textwidth}
\centering
\begin{tikzpicture}[
    scale=0.85,
    every node/.style={scale=0.7}
]
    \pgfmathsetmacro{\n}{3}
    \pgfmathsetmacro{\nm}{2}
    \pgfmathsetmacro{\tsize}{\tlength*\n}
    
    \node (ha) at (0,0) {};
    
    \pgfmathsetmacro{\startX}{iseven(\n) ? (\n/2)*(\tlength) : (\nm/2)*\tlength + \qzlength }
    \pgfmathsetmacro{\centerLength}{iseven(\n) ? \qzlength : \qolength }
    
    \node (ps) at (0,0) {};
    \node (pe) at (2.6,0) {};
    
    \foreach \x in {0,...,\nm} {
        \draw ($(\x*\tlength,0)$) rectangle ++ (\qzlength,\rheight) node[midway] {$q_0$};
        \draw ($(\x*\tlength+\qzlength,0)$) rectangle ++ (\qolength,\rheight) node[midway] {$q_1$};

        \ifthenelse{\x=0}
          {\def\mycolor{red}}
          {\def\mycolor{blue}}
        \ifthenelse{\x=0 \OR \x=\nm}
          {\def\rcolor{red}}
          {\def\rcolor{blue}}
        \draw [pen colour={\mycolor}, decorate, orange, decoration = {calligraphic brace,mirror}] ($(\x*\tlength,-0.5*\rheight)$) -- ($(\x*\tlength+\qzlength,-0.5*\rheight)$)
        node[midway,shift={(0,{-0.7*\rheight})},black] {};

        \draw [pen colour={\rcolor}, decorate, orange, decoration = {calligraphic brace}] ($(\x*\tlength+\qzlength,1.5*\rheight)$) -- ($(\x*\tlength+\tlength,1.5*\rheight)$)
        node[midway,shift={(0,{-0.7*\rheight})},black] {};
    }

    \draw ($(\n*\tlength,0)$) rectangle ++ (\qzlength,\rheight) node[midway] {$q_0$};
    \draw [pen colour={red}, decorate, orange, decoration = {calligraphic brace,mirror}] 
    ($(\n*\tlength,-0.5*\rheight)$) -- ($(\n*\tlength+\qzlength,-0.5*\rheight)$)
    node[midway,shift={(0,{-0.7*\rheight})},black] {};

\end{tikzpicture}
\caption{Case 1. Both $c_1,\,c_2$ are internal centers of the same palindromic run. Non-internal centers are highlighted with a red curly brace.}
\end{subfigure}

\begin{subfigure}{\textwidth}
\centering
\begin{tikzpicture}[scale=0.85, every node/.style={scale=0.7}]
    \pgfmathsetmacro{\n}{3}
    \pgfmathsetmacro{\nm}{2}
    \pgfmathsetmacro{\tsize}{\tlength*\n}
    
    \node (ha) at (0,0) {};
    
    \pgfmathsetmacro{\startX}{iseven(\n) ? (\n/2)*(\tlength) : (\nm/2)*\tlength + \qzlength }
    \pgfmathsetmacro{\centerLength}{iseven(\n) ? \qzlength : \qolength }
    
    \draw (0,0) rectangle ++ ($(\n*\tlength+\qzlength,\rheight)$) node[midway] {};

    \draw [pen colour={blue}, decorate, orange, decoration = {calligraphic brace,mirror}] 
    ($(\tlength,-0.5*\rheight)$) -- ($(\n*\tlength,-0.5*\rheight)$)
    node[midway,shift={(0,{-0.7*\rheight})},black] {$\Runc2$};

    \draw [pen colour={red}, decorate, orange, decoration = {calligraphic brace}] 
    ($(2*\tlength,1.5*\rheight)$) -- ($(\n*\tlength-1,1.5*\rheight)$)
    node[midway,shift={(0,{0.7*\rheight})},black] {$\Runc1$};
\end{tikzpicture}
\caption{Case 2. One palindromic run is contained inside the other. For example, consider the string $S=xxxyxxxyxxx$, where $x$ is some arbitrary palindrome.}
\end{subfigure}

\begin{subfigure}{\textwidth}
\centering
\begin{tikzpicture}[scale=0.85, every node/.style={scale=0.7}]
    \pgfmathsetmacro{\n}{3}
    \pgfmathsetmacro{\nm}{2}
    \pgfmathsetmacro{\tsize}{\tlength*\n}
    
    \node (ha) at (0,0) {};
    
    \pgfmathsetmacro{\startX}{iseven(\n) ? (\n/2)*(\tlength) : (\nm/2)*\tlength + \qzlength }
    \pgfmathsetmacro{\centerLength}{iseven(\n) ? \qzlength : \qolength }
    
    \draw (0,0) rectangle ++ ($(\n*\tlength+\qzlength,\rheight)$) node[midway] {};

    \draw [pen colour={blue}, decorate, orange, decoration = {calligraphic brace,mirror}] 
    ($(\tlength+1,-0.5*\rheight)$) -- ($(\n*\tlength,-0.5*\rheight)$)
    node[midway,shift={(0,{-0.7*\rheight})},black] {$\Runc2$};

    \draw [pen colour={red}, decorate, orange, decoration = {calligraphic brace}] 
    ($(1,1.5*\rheight)$) -- ($(\n*\tlength-1-\tlength,1.5*\rheight)$)
    node[midway,shift={(0,{0.7*\rheight})},black] {$\Runc1$};
\end{tikzpicture}
\caption{Case 3. The palindromic runs possibly overlap, but do not contain one another. As an example, consider the string $S=xzxzxyxyx$, where $x,\,y,\,z$ are replaceable by arbitrary palindromes.}
\end{subfigure}
    \caption{Demonstration of the different cases in~\cref{lem:per-len}.}
    \label{fig:per-len-demo}
\end{figure}

A visual demonstration of the different cases of the proof can be found in~\cref{ap:sec:examples} as~\cref{fig:per-len-demo}.

\begin{proof}[Proof for~\cref{lem:layb}]\label{pr:lem:layb}
    The proof presents the construction of $\I$ and the corresponding query procedure, followed by a correctness proof.

    \textbf{Construction.} We construct a hierarchy of bitvectors $\I$, where each bitvector in the hierarchy is preprocessed for \mathttt{rank/select} queries. \\
    The first bitvector $\I[1]$ has $ 2n $ bits, the second bitvector $\I[2]$ has $ n $ bits, and each new level is half the size of the preceding; the total space is $\Ohs{n}$ bits. Since the largest possible period is $n/2$, the number of layers is $\lceil \log (n/2+1)\rceil$.
    
    The bitvector at layer $ m $ (i.e., $\I[m]$) stores information for periodic centers whose period falls within the interval $ [2^{m-1},2^{m}) $. In the bitvectors, a set bit (`1') indicates the first internal center of a palindromic run, and the other bits are unset (`0'). Accessing the bit corresponding to center $c$ at level $m$ requires scaling. We denote the scaling operation as $\operatorname{scl}$, defined as $\scl{c}{m}\coloneq \shr{2c}{m-1}$.

    Formally, given a palindromic run $\Run$ with period $p$:
    \begin{enumerate}
        \item Determine the layer $m$ that satisfies $2^{m-1}\le p<2^m$.
        \item Denote the set of internal centers of $\Run$ as $C_{\mathrm{internal}}$.
        \item Set the first internal center of $\Run$; $c_1\coloneq \min\{C_{\mathrm{internal}}\}$.
        \item Set $c_m^{(1)}\gets \scl{c_1}{m}$.
        \item Assign $\I[m][c_m^{(1)}]\gets 1$.
    \end{enumerate}
    Given a hierarchy $\I$ where all bitvectors initially contain only zeroes, we perform the above procedure for every palindromic run in $S$. Then, we preprocess all bitvectors in the hierarchy to support \mathttt{rank/select} queries in constant time. The resulting hierarchy $\I$ is our data structure.

    \textbf{Query.} The query procedure computes the first internal center $c_1$ of $\Run_c$ given a periodic center $c$. The procedure consists of five steps:
    \begin{enumerate}
        \item Retrieve the period of $\Run_c$ from $\CP$; $p_c\gets \CP[2c]$.
        \item Determine the appropriate layer $m$ that satisfies $2^{m-1}\le p_c<2^m$.
        \item Compute the scaled index for center $c$; $c_m\gets \scl{c}{m}$.
        \item Perform a predecessor query on $\I[m]$ at position $c_m$ via $\mathttt{rank/select}$ queries to get an approximation of the required center:
        \[
            \apc\gets \operatorname{select}_1(\I[m],\;\operatorname{rank}_1(\I[m],\;c_m)).
        \]
        \item Compute the resulting index $c_1$. The resulting index satisfies:
        \[
        (i)\quad \shl{\apc}{m-1} \le 2c_1< \shl{\apc+1}{m-1},\qquad (ii)\quad 2(c-c_1)= 0\quad \text{(mod $p_c$)}.
        \]
        Since $2^{m-1}\le p_c$, at most one possible center satisfies the above conditions, as two periodic centers are separated by $p_c\ge 2^{m-1}$ array positions. Therefore, given that $\apc=\scl{c_1}{m}$, the correct result will always be returned. 
    \end{enumerate}

    \textbf{Correctness.} The correctness of the data structure relies on the smoothness of the center-period array $\CP$. We denote $\Runi1\coloneq \Run_{c}$.
    
    For the query to fail, the bitvector $\I[m]$ must identify an invalid internal center $c_1'>c_1$ as the first internal center of $\Runi1$. Denote $\Runi2\coloneq \Run_{c_1'}$. Since $c_1'$ is indicated in level $m$, it follows that $\len{\rooti2}\in[2^{m-1},2^m)$.
    
    The collision implies that the scaled indices satisfy \\ $\scl{c_1}{m} < \scl{c_1'}{m} \le \scl{c}{m}$. Because $2^{m-1}\le \len{\rooti1}<2^m$, a center $\hat{c}$ exists such that $2\abs{\hat{c}-c_1'}<2^{m-1}$. However, $\min(\len{\rooti1},\,\len{\rooti2})\ge 2^{m-1}$, which implies $\min(\CP[2\hat{c}],\,\CP[2c_1'])\ge 2^{m-1}>2\abs{\hat{c}-c_1'}$,
    violating the smoothness condition of $\CP$, hence contradicting~\cref{lem:per-len}.
\end{proof}

A detailed example of the construction and the query procedure can be found in~\cref{ap:ex:layb}. 

\subsection{Demonstration of the layered bitvector}\label{ap:sec:examples}

This section offers a detailed demonstration of the layered bitvector structure.

\begin{figure}[ht]
\centering
\begin{tikzpicture}[x=0.35cm,y=0.8cm,>=stealth]

\def\legendx{26}
\def\legendy{-3.4}

\begin{scope}[shift={(\legendx,\legendy)}]

  \draw[rounded corners] (0,0) rectangle (9,-1.8);

  \draw[fill=black,draw=none] (0.4,-0.4) circle (1.5pt);
  \node[anchor=west] at (0.8,-0.4) {\scriptsize repeat center};

  \draw[fill=blue,draw=none] (0.4,-0.9) circle (1.5pt);
  \node[anchor=west] at (0.8,-0.9) {\scriptsize internal center};

  \draw[red,thick] (0.4,-1.4) circle (3pt);
  \node[anchor=west] at (0.8,-1.4) {\scriptsize first internal center};
\end{scope}

\def\scriptSizeStart{6}
\def\initIndexL{3}
\def\initIndexS{2.5}
\def\initIndexH{2.5}
\def\textH{0.3}
\def\boxW{0.6}

\node[above] at (16,\initIndexH+1.4) {The bitvectors hierarchy $\mathbf{I}$ for $S=\mathttt{aaaaabaaaaabaaaaabaaaaabaaaaabcbcb}$.};

\node[left] at (\initIndexL,\initIndexH+0.3) {$\mathbf{I}[1]$};

\foreach \i in {1,...,18}{
  \draw (\i+\initIndexS,\initIndexH) rectangle ++(1,0.6);
  \node at (\i+\initIndexL,\initIndexH+\textH) {\tiny 0};
}
\foreach \i in {19,19.5,20,20.5,21}{
  \node at (\i+\initIndexL,\initIndexH+\textH) {\tiny .};
}
\foreach \i in {22,...,30}{
  \draw (\i+\initIndexS,\initIndexH) rectangle ++(1,0.6);
  \node at (\i+\initIndexL,\initIndexH+\textH) {\tiny 0};
}

\foreach \b in {3,15} {
    \draw[fill=black!10] (\b+2.5,\initIndexH) rectangle ++(1,0.6);
    \node at (\b+\initIndexL,\initIndexH+\textH) {\tiny 1};
}

\foreach \i in {1,5,9,13,17}{
  \node[above] at (\i+\initIndexL,\initIndexH+0.6) {\scriptsize \i};
}
\foreach \i in {60,64,68}{
  \node[above] at (\i+\initIndexL-38,\initIndexH+0.6) {\scriptsize \i};
}

\foreach \i/\ch in {
  1/a,2/a,3/a,4/a,5/a,6/b,
  7/a,8/a,9/a,10/a,11/a,12/b,
  13/a,14/a,15/a,16/a,17/a,18/b,
  19/a,20/a,21/a,22/a,23/a,24/b,
  25/a,26/a,27/a,28/a,29/a,30/b,
  31/c,32/b,33/c,34/b
}{
  \draw (\i,0) rectangle ++(1,0.8);
  \node at (\i+0.5,0.4) {\small\mathttt{\ch}};
}

\foreach \r in {3, 9, 15, 21, 27} {
    \foreach \c in {0,1,9,10}{
      \draw[fill=black,draw=none] (\r-2+\c/2,0.9) circle (1.2pt);
    }
    \foreach \c in {2,3,4,5,6,7,8}{
      \draw[fill=blue,draw=none] (\r-2+\c/2,0.9) circle (1.2pt);
    }

    \draw[red] (\r-1,0.9) circle (2.2pt);

}

\foreach \c in {1.5, 7.5, 13.5, 19.5, 25.5} {
    \node[above,red] at (\c+0.5,0.9) {\scriptsize $c=\c$};
}

\def\c{1.5}
\pgfmathsetmacro{\cc}{int(\c*2)}
\draw[->, thick]
  (\c+0.5,1.2) .. controls +(0,0.6) and +(0,-1) ..
  (\cc+\initIndexL,\initIndexH-0.35);

\node[below left] at (\cc+\initIndexL,\initIndexH-0.35) {\scriptsize $\scl{\c}{1}=\cc$};

\def\c{7.5}
\pgfmathsetmacro{\cc}{int(\c*2)}
\draw[->, thick]
  (\c+0.5,1.2) .. controls +(1,0.5) and +(0, -0.5) ..
  (\cc+\initIndexL,\initIndexH-0.35);
\node[below left] at (\cc+\initIndexL-3,\initIndexH-0.3) {\scriptsize $\scl{\c}{1}=\cc$};

\draw[->]
  (13.5,1.2) .. controls +(0,0.5) and +(0,-0.5) ..
  (19+\initIndexL,\initIndexH-0.35);
\draw[->]
  (19.5,1.2) .. controls +(0,0.5) and +(0,-0.5) ..
  (20+\initIndexL,\initIndexH-0.35);
\draw[->]
  (25.5,1.2) .. controls +(0,0.5) and +(0,-0.5) ..
  (21+\initIndexL,\initIndexH-0.35);

\foreach \c in {3,6,27,30,31,33,34}{
  \draw[fill=black,draw=none] (\c+0.5,-0.1) circle (1.5pt);
}

\foreach \c in {9,12,15,18,21,24,32}{
  \draw[fill=blue,draw=none] (\c+0.5,-0.1) circle (1.5pt);
}

\foreach \c in {9, 32} {
    \draw[red] (\c+0.5,-0.1) circle (3pt);
}

\foreach \c in {9, 32} {
    \node[below,red] at (\c+0.5,-0.4) {\scriptsize $c=\c$};
}
\foreach \c in {12,15,18,21,24} {
    \node[below,black!30!blue] at (\c+0.5,-0.4) {\scriptsize $c=\c$};
}

\def\initIndexH{-2.6}

\draw[->,thick]
  (9.5,-0.7) .. controls +(0,-1.3) and +(0,1.3) ..
  (5+\initIndexL,\initIndexH+0.8);
\node[above left] at (5+\initIndexL,\initIndexH+0.8) {\scriptsize $\scl{9}{3}=5$};

\draw[->,thin]
  (24.5,-0.7) .. controls +(0,-1.3) and +(1,1) ..
  (12+\initIndexL,\initIndexH+0.8);
\node[above right] at (12+\initIndexL,\initIndexH+0.6) {\scriptsize $\scl{24}{3}=12$};

\node[left] at (\initIndexL,\initIndexH+0.3) {$\mathbf{I}[3]$};

\foreach \i in {1,...,17}{
  \draw (\i+\initIndexS,\initIndexH) rectangle ++(1,0.6);
  \node at (\i+\initIndexL,\initIndexH+\textH) {\tiny 0};
}

\draw[fill=black!10] (7.5,\initIndexH) rectangle ++(1,0.6);
\node at (5+\initIndexL,\initIndexH+\textH) {\tiny 1};

\foreach \i in {1,5,9,13,17}{
  \node[below] at (\i+\initIndexL,\initIndexH) {\scriptsize \i};
}

\edef\initIndexH{\number\numexpr\initIndexH - 1.25}
\node[left] at (\initIndexL,\initIndexH+0.3) {$\mathbf{I}[2]$};
\foreach \i in {1,...,13}{
  \draw (\i+\initIndexS,\initIndexH) rectangle ++(1,0.6);
  \node at (\i+\initIndexL,\initIndexH+\textH) {\tiny 0};
}
\foreach \i in {14,14.5,15}{
  \node at (\i+\initIndexL,\initIndexH+\textH) {\tiny .};
}
\foreach \i in {16,...,21}{
  \draw (\i+\initIndexS,\initIndexH) rectangle ++(1,0.6);
  \node at (\i+\initIndexL,\initIndexH+\textH) {\tiny 0};
}
\foreach \i in {0,...,3}{
  \pgfmathsetmacro{\l}{int(\i*4+1)}
  \node[below] at (\l+\initIndexL,\initIndexH) {\scriptsize \l};
}
\foreach \i/\l in {17/30,21/34}{
  \node[below] at (\i+\initIndexL,\initIndexH) {\scriptsize \l};
}

\draw[->,thick]
  (32.5,-0.7) .. controls +(0,-1.3) and +(0,1.3) ..
  (19+\initIndexL,\initIndexH+0.8);
\node[above right] at (19.5+\initIndexL,\initIndexH+0.8) {\scriptsize $\scl{32}{2}=32$};

\foreach \i in {19} {
    \draw[fill=black!10] (\i+2.5,\initIndexH) rectangle ++(1,0.6);
    \node at (\i+\initIndexL,\initIndexH+\textH) {\tiny 1};
}

\edef\initIndexH{\number\numexpr\initIndexH - 1.4}
\node[left] at (\initIndexL,\initIndexH+0.3) {$\mathbf{I}[4]$};
\foreach \i in {1,...,8}{
  \draw (\i+\initIndexS,\initIndexH) rectangle ++(1,0.6);
  \node at (\i+\initIndexL,\initIndexH+\textH) {\tiny 0};
}
\foreach \i in {1,5}{
  \node[below] at (\i+\initIndexL,\initIndexH) {\scriptsize \i};
}

\edef\initIndexS{\number\numexpr\initIndexS + 15}
\edef\initIndexL{\number\numexpr\initIndexS + 0.5}

\node[left] at (\initIndexL,\initIndexH+0.3) {$\mathbf{I}[5]$};
\foreach \i in {1,...,4}{
  \draw (\i+\initIndexS,\initIndexH) rectangle ++(1,0.6);
  \node at (\i+\initIndexL,\initIndexH+\textH) {\tiny 0};
}
\foreach \i in {1}{
  \node[below] at (\i+\initIndexL,\initIndexH) {\scriptsize \i};
}

\end{tikzpicture}
\caption{The relation between internal centers of runs in $S$ to the layered bitvector $\mathbf{I}$. Above are the repeat centers of all palindromic runs with period $1$, i.e., repetitions $\mathttt{aaaaa}$. Below are the repeat centers of $\Run_{15}$ and $\Run_{32}$. For clarity, each center is classified with respect to a specific palindromic run.}
\label{fig:layered-bitvector}
\end{figure}

\begin{example}[Example for~\cref{lem:layb}]\label{ap:ex:layb}

    Let $S=\mathttt{aaaaabaaaaabaaaaabaaaaabaaaaabcbcb}$ be a string. The string has seven distinct palindromic runs. The longest run, $\Run=\Run_{15}$, is described as
    \[
        \Run=[1\dd 29],\qquad \root=q_0q_1,\qquad q_0=\mathttt{aaaaa},\quad q_1=\mathttt{b}.
    \]
    
    Five other palindromic runs are the different instances of $q_0$ in $S$, and have period one, the following root:
    \[
        \root'=q'_0q'_1,\qquad q'_0=\emptystring,\quad q'_1=\mathttt{a}.
    \]

    The last palindromic run is the periodic suffix of the string, and is described by:
    \[
        \Runi2=[30\dd 34],\qquad \rooti2=q_0^{(2)}q_1^{(2)},\qquad q_0^{(2)}=\mathttt{b},\quad q_1^{(2)}=\mathttt{c}
    \]
    
    As previously mentioned, this data structure intends to locate the first internal center of a given palindromic run, given any internal center of that specific run. Constructing a data structure for locating the last internal center is analogous to this construction.

    In this example, the input string length is $n=34$. Since the data structure receives centers as inputs, the first layer is of size $2n=68$. The largest possible period in a string is $n/2=17$, and therefore the number of layers in the bitvector is $\lceil \log (n/2+1)\rceil=\lceil\log 18 \rceil=5$. The details of the layers are in the following table:
    \begin{center}
    \renewcommand{\arraystretch}{1.3}
    \setlength{\tabcolsep}{10pt}
    \begin{tabular}{c | l | c  }
        \textbf{Layer} & \textbf{Period range} & \textbf{Number of bits} \\
        \hline
        $\I[1]$ & $[1, 2)$     & 68  \\
        $\I[2]$ & $[2, 4)$     & 34  \\
        $\I[3]$ & $[4, 8)$     & 17  \\
        $\I[4]$ & $[8, 16)$    & 8   \\
        $\I[5]$ & $[16, 17]$   & 4   
    \end{tabular}
    \end{center}

    In our example, $\Run$ has period $6$ and belongs in layer three. The other runs have period one and belong in the first layer, and the last run has period $2$ and belongs in the second layer.

    Recall the scaling operation $\operatorname{scl}$:
    \[
    \scl{c}{m}\coloneq \shr{2c}{m-1}=((2c - 1)\gg (m-1)) + 1.
    \]

    We begin with the five palindromic runs with period one. Each of the five runs has eleven repeat centers, hence five different internal centers. Consider, for example, $\Run_3=[1\dd 5]$. The repeat centers of $\Run_3$ are $C_{\text{rep}}(\Run_3)=\{0.5,1,1.5,\dots,5.5\}$, of which, $\{1.5,2,\dots,4,4.5\}$ are its internal centers, and $1.5$ is the first of them. Altogether, the first internal centers of all of those runs are $\{1.5,7.5,13.5,19.5,25.5\}$. At layer one the scaling operation satisfies $\scl{c}{1}=2c$. Therefore, the indices that should be set to `1' in $\I[1]$ are $\{3, 15, 27, 39, 51\}$, and as there are no other runs with period one, the resulting bitvector has `1' in the specified indices and `0' in the rest.

    Proceeding to the next layer, we recall that we only store information about internal centers. Since $\Runi2$ has only five repeat centers, its only internal center (and consequently, the first internal center) is $32$. The scaling at layer two satisfies $\scl{c}{2}=c$, hence we only set $\I[2][32]\gets 1$, and the rest of the bits are unset.
    
    Lastly, we update the bits corresponding to $\Run$. The repeat centers of $\Run$ are $\{3,6,9,\dots,27,30\}$. Of those centers, $\{9,12,15,18,21,24\}$ are internal centers. After scaling to level three, the corresponding indices in $\I[3]$ are $\{5, 6, 8, 9, 11, 12\}$. Like before, we only set the bit corresponding to the first internal center, i.e., $9$. And since $\scl{9}{3}=5$, the third bitvector with seventeen bits is $\I[3]=\mathttt{00001000000000000}$.

    Lastly, we demonstrate how to find the first internal center, given the periodic center $c=18$ as input:
    \begin{enumerate}
        \item We compute the period $p\coloneq |\root_c|$ from the modified center-period array. 
        \[
            p\gets \CP[2c]=\CP[36]=6.
        \]
        \item We compute the target layer $m$, satisfying $2^{m-1}\le |\root_c|<2^{m}$. The result is $m=3$.
        \item We compute the scaled index:
        \[
            c_m \gets \scl{c}{m}=\scl{18}{3}=\shr{36}{2}=9.
        \]
        \item We perform predecessor query on $\I[m]$ using \mathttt{rank/select} queries:

        \begin{align*}
            x\gets \operatorname{rank}_1(\I[m],\,c_m)=\operatorname{rank}_1(\I[3],\,9)=1\\
            \apc\gets \operatorname{select}_1(\I[m],\,x)=\operatorname{select}_1(\I[3],\,1)=5
        \end{align*}
        \item Given $\apc$, we find the original center that satisfies:
        \[
        (i)\quad \shl{\apc}{m-1} \le 2c_1< \shl{\apc+1}{m-1},\qquad (ii)\quad 2(c-c_1)= 0\quad \text{(mod $p$)}.
        \]
        And when plugging in the values:
        \[
        17 \le 2c_1< 21,\qquad \quad 36-2c_1= 0\quad \text{(mod $6$)}.
        \]
        The only possible value of $c_1$ to satisfy this modular equality is $c_1=9$, which is the center we sought - the first internal center of $\Run$.
    \end{enumerate}

    A visualization of this data structure can be found in~\cref{fig:layered-bitvector}.
\end{example}

\section{Future Work}

While our data structure achieves asymptotically optimal space and constant access time, several questions remain. First and foremost: a full and formal construction algorithm running in linear time is a natural next step, potentially using $\Oh{n}$ bits of working space. Another open problem is determining the exact entropy of the Manacher array. Finally, investigating the adequacy of our techniques for other combinatorial structures in string processing may further advance compressed indexing and retrieval.

\newpage

\bibliography{paper}
\newpage

\appendix

\section{Sketching the Construction}\label{ap:sec:const-sketch}

In this appendix section, we sketch a linear-space, $\tilde{\mathcal{O}}(n)$-time algorithm to construct the compressed Manacher array.

To construct the data structure correctly, we need to distinguish between periodic and non-periodic centers. We proved in~\cref{lem:no-ext-pal} that if $c$ is a periodic center, $\Pal_c$ is a prefix (or suffix) of $\Run_c$. Therefore, given a string $S$, we compute all maximal repetitions in linear time~\cite{KK:99}, which yields an array $\mathcal R(S)$. For every maximal repetition, we store its information in both endpoints, i.e., if $S[i\dd j]$ is a maximal repetition, we store the information in $\mathcal R(S)[i]$ and $\mathcal R(S)[j]$, increasing the space at most by a factor of two. At most $\Oh{\log n}$ maximal repetitions can start or end in an index~(Corollary 1 in \cite{Crochemore2009}).

Having computed the array $\mathcal R(S)$, we compute the sparse Manacher array $\tilde{\mathsf A}$, initially equal to the original Manacher array $\mathsf A$. For each subpalindrome $\Pal_c=[i\dd j]$, we inspect its endpoints $i$ and $j$ and test against at most $\Ohs{\log n}$ maximal repetitions stored at its endpoints whether it is periodic. If so, we store the necessary information in the enriched center-period array $\tilde{\mathsf L}$, and set its corresponding value in $\mathsf A$ to zero. As discussed earlier, the length of a periodic palindrome is $k|\root|+|q_0|$, and therefore we compute $|q_0|=|\Pal_c|\bmod |\root|$.

Lastly, we compute $\mathsf I$. For each periodic center $c$, we retrieve its $\PPD$ and check if $|\Pal_c|=2|\root|+|q_0|$. If so, then $c$ is the first internal center, in which case we set the corresponding bit in $\mathbf I$ to one. After processing all periodic centers, we preprocess the bit vectors in $\mathbf I$ for $\operatorname{rank}/\operatorname{select}$ queries, in overall linear time. Finally, we construct $\mathsf{I}$ by building a symmetric structure $\mathbf I_{\text{last}}$ to find the last internal center.

At this point, all components of the data structure are ready. The entire construction can be carried out in linear space and $\Ohs{n\log n}$ time. An $\Ohs{n}$-time construction appears attainable with moderate refinements

\end{document}